\documentclass[seceq]{ptptex}
\usepackage{type1cm, 
hyperref}

\newcommand{\p}{\partial}
\newcommand{\G}{\Gamma}

\newcommand{\ti}{\tilde}
\newcommand{\al}{\alpha'}

\newcommand{\f}{\frac}

\preprintnumber[3cm]{
UT-06-**\\ hep-th/0612***}

\title{Winding String Dynamics \\
in a Time-Dependent Beta Deformed Background}

\author{Ta-Sheng \textsc{Tai}\footnote{E-mail:
tasheng@hep-th.phys.s.u-tokyo.ac.jp}}

\inst{Department of Physics, University of Tokyo, Tokyo 113-0033, Japan}

\abst{%
We study string theory on the analytically continued 
$\beta$ deformed background proposed in hep-th/0509036. 
This non-static model provides a solvable conformal 
field theory which describes time-dependent twisted string dynamics. 
With the mini-superspace approach, 
we examine disk one-point correlators of D-branes and 
compute winding string pair production rate. 
We find that these results 
are consistent with the CFT computation. 
}

\begin{document}
\maketitle

\section{Introduction and summary}

Winding string dynamics in Lorentzian orbifolds have recently been widely studied 
(see 
Refs.~1) -- 9) and references therein). 
Because string theory in that context 
takes the form of a solvable conformal 
field theory, these time-dependent backgrounds provide interesting toy models 
in which 
both the mini-superspace approximation and the worldsheet theory can be examined.

In this note, we study one exactly solvable non-static string background 
proposed in Ref.~10). This background is obtained by applying Lorentzian analytical continuation to 
the static $\beta$ deformed background, which has a metric similar to that of a flux-brane. 
As was first shown 
in Refs.~11) and 12), the string sigma model associated with 
the $\beta$ deformed background is equivalent to an orbifold after the 
so-called TsT transformation is carried out. 
Following Ref.~13), the prescription to generate this kind of metric 
consists of the application of 
an $SL(2,R)$ transformation to the parameter $\tau$ as 
\begin{eqnarray}
\tau=B_{21}+i\sqrt{g}~\rightarrow ~\tau'=\frac{\tau}{1+b \tau} ~
\label{tau}
\end{eqnarray}
for any string background with two $U(1)$ 
symmetries realized geometrically, i.e. a 
two-torus parameterized by, say, $\varphi_1$ and $\varphi_2$. 
Note that $B_{21}$ is the $B$-field along $\varphi_1$ and $\varphi_2$, 
while $\sqrt{g}$ denotes the volume of the two torus under 
consideration. For instance, given a flat ${\mathbb R}^4$ metric 
$ds^2 =dr_1^2 + dr_2^2 + {r_1^2}d\varphi_1^2 + {r_2^2}d\varphi_2^2$, 
after the transformation given in Eq.~$\eqref{tau}$, we have 
\begin{eqnarray}
&&ds^2 =dr_1^2 + dr_2^2 + \f{r_1^2}{1+ b^2 r_1^2 r_2^2}d\varphi_1^2 + \f{r_2^2}{1 + b^2 r_1^2 r_2^2}d\varphi_2^2 ~,\nonumber\\ 
&&B_{\varphi_1 \varphi_2}= \f{-br_1^2 r_2^2}{1 + b^2 r_1^2 r_2^2} ~,~~~~~
e^{2(\Phi-\Phi_0)}=\f{1}{1+b^2 r_1^2 r_2^2 } ~,
\label{melv}
\end{eqnarray}
where we choose 
$b>0$ and $\varphi_i\sim \varphi_i + 2\pi$, $0\le r_i\le\infty$ $(i=1,2)$. 
Then, applying the Wick rotation $r_1\rightarrow it$, $\varphi_1\rightarrow i\theta$, 
$b \rightarrow ib$, Eq.~$\eqref{melv}$ becomes 
\begin{eqnarray}
&&ds^2 = -dt^2 + dr^2 + \f{t^2}{1+ b^2 t^2 r^2}d\theta^2 + 
\f{r^2}{1 + b^2 t^2 r^2}d\varphi^2 ~,\nonumber\\ 
&&B_{\theta \varphi}= \f{-bt^2 r^2}{1 + b^2 t^2 r^2} ~,~~~~~
e^{2(\Phi-\Phi_0)}=\f{1}{1+b^2 t^2 r^2 } ~,
\label{me2}
\end{eqnarray}
where $-\infty<t,\theta<\infty$, $0\le r\le\infty$ and $\varphi\sim\varphi+ 2\pi$. 
Though Eq.~$\eqref{me2}$ is generated from 
one half of ${\mathbb R}^{1,3}$, 
it is possible to extend to the entire ${\mathbb R}^{1,3}$ and thereby 
to obtain a quantizable Lorentzian string 
orbifold through a TsT operation, as in Refs.~6), 7), and 14). 
As is pointed out in Ref.~10), the metric $\eqref{me2}$ possesses an 
interesting time dependence. 
When $t\sim0$, this geometry 
has the same form as ${\mathbb R}^{1,1}\times{\mathbb C}$. Then, 
as time elapses, a bubble appears near $r=0$ of 
${\mathbb C}$ whose size is approximately 
$r=\f{1}{b|t|}$. Further, for $t\sim\infty$,  the metric $\eqref{me2}$ becomes 
a slender cone ${\mathbb C}/{\mathbb Z}_{\infty}$ times 
a Misner space 
orbifolded by a boost $\triangle=2\pi\al b$, i.e. 
$[{\mathbb R}^{1,1}/{\mathbb Z}]_{\triangle=2\pi \al b}$.

On the other hand, under the mini-superspace approximation, 
in which the string worldsheet reduces to a point particle, the 
string dynamics in the background $\eqref{me2}$ are generally 
described by a second-order ordinary 
differential equation 
\begin{align}
\ddot{{\psi}}(t)+\omega(t)^2 {\psi}(t)=0 ~,
\label{11}
\end{align}
up to a necessary change of variable, where the dot denotes differentiation w.r.t. time. 
This can be understood as a result of 
the on-shell constraint for a given string state $\psi$. 
Because the metric $\eqref{me2}$ is 
non-static, any quantum field propagating within it 
gives rise to particle production. 
Therefore, one can solve Eq.~$\eqref{11}$ to obtain the 
Bogoliubov coefficients which 
encode the production rate. We found that 
the production rate consists of two parts 
and is analogous to the case of charged 
particle pair creation in the Rindler space;$^{17)}$ 
that is, one part, which is induced by the background $B$-field in 
$\eqref{me2}$, corresponds to the 
usual Schwinger effect, whereas the other, which is due to the Milne metric, i.e. 
$b=0$ in $\eqref{me2}$, corresponds to the gravitational Unruh effect of Rindler dynamics. 
Moreover, 
by means of the DBI action, it is also possible to obtain the classical 
one-point correlator by computing the overlap of $\psi(t)$ and the D-brane worldvolume.

In addition, from the exact string spectrum on the background $\eqref{me2}$ after TsT, 
we can compute 
the torus and cylinder amplitudes, which facilitate a non-trivial comparison with 
the mini-superspace analysis. Though there is consistency between the results on 
disk one-point correlators, only the (anti-) holomorphic quasi-zero 
mode part of the torus amplitude 
is reproduced by the Bogoliubov coefficients of wave functions. 
Moreover, the double-pole structure of 
the torus amplitude, which arises from the product of the left and right mover 
of closed strings, has no imaginary part after the 
moduli integral is carried out. By considering the optical theorem, this seems to 
contradict the observation concerning string production made previously. 
However, there is, in fact, no contradiction. The apparent contradiction is 
simply due to 
the difference in the choice of the $in$ and $out$ vacua. 
In CFT, the $in$ and $out$ vacua coincide, because 
the physical descriptions are the same at $t=\pm\infty$, 
as seen from the metric $\eqref{me2}$, 
while in the mini-superspace approach, we have 
chosen them to be at $t=0$ and $t=\infty$, respectively.

The rest of this note is organized as follows. In $\S$2, we derive scalar wave 
functions. There, string production and disk 
one-point correlators are 
examined. In $\S$3, we compare previous results with those obtained from the CFT 
calculation.

\section{Mini-superspace analysis}

In the mini-superspace approximation, the worldsheet spectrum is truncated at zero-modes, 
and the closed string Hamiltonian $L_0 +\ti L_0$ reduces to the effective Laplacian 
\begin{align}
\triangle_0=
\frac{1}{e^{-2\Phi}\sqrt{-{\text {det}}G}}\p_{\mu}
\Big(e^{-2\Phi}\sqrt{-{\text {det}}G}~G^{\mu \nu}\p_{\nu}\Big) ~.
\label{L}
\end{align}
Plugging the metric $\eqref{me2}$ into this Laplacian, we obtain a scalar wave equation. 
Here, we write 
down its solutions and study the winding string pair production process. Then, we also 
derive classical disk one-point correlators of probe D-branes. 

\subsection{Wave functions}

To begin, let us derive the full string spectrum and then consider the 
zero-slope 
limit to truncate at zero-modes. This facilitates a comparison with the spectrum 
derived from the wave mechanical approach sketched above. The sigma model associated 
with closed strings propagating in $\eqref{me2}$ is 
\begin{eqnarray}
S&=&\f{-1}{4\pi \al}\int 
\Big[-dt \wedge^\ast dt  + dr \wedge^\ast dr + \f{1}{1+ b^2 t^2 r^2}(t^2d\theta \wedge^\ast d\theta)\nonumber\\
&&\hspace*{1.6cm}+ \f{1}{1+ b^2 t^2 r^2}(r^2d\varphi \wedge^\ast d\varphi)
- \f{bt^2 r^2}{1 + b^2 t^2 r^2}(d\theta \wedge d\varphi)\Big] ~.
\label{tsig}
\end{eqnarray}
Following Ref.~10), we first compactify $\theta$ to some finite radius 
and then T-dualize 
it to $\tilde \theta$, 
shifting $\varphi$ as $\varphi'= \varphi + b\tilde\theta$. 
Next, we T-dualize 
again from $\tilde\theta$ to $\theta'$, 
and finally decompactify $\theta$. This process 
yields a free sigma model in terms of $X^{\pm}=\f{1}{\sqrt{2}}te^{\pm \theta'}$ 
and 
$X=re^{i\varphi'}$, which satisfy%
\footnote{The same operation can be carried out 
for the space-like counterpart, where $X^+ X^-<0$.} 
\begin{eqnarray}
X^{\pm}(\tau,\sigma+2\pi)&=e^{\pm 2\pi\nu}X^{\pm}(\tau,\sigma) ~,~~\nu=\beta{(J_L+J_R)} ~,\nonumber\\
X(\tau,\sigma+2\pi)&=e^{-2\pi i\mu}X(\tau,\sigma) ~,~~ 
\mu=-\beta{({\cal J}_L+ {\cal J}_R)} ~,
\label{pe} 
\end{eqnarray}
where $\beta=\al b$. 
Note that ${\cal J}_{L} + {\cal J}_{R}$ $(J_{L} + J_{R})$ is the 
boost (rotation) generator of 
the $X^{\pm}$ $(X)$ plane. 
Upon embedding this in the 26-dimensional bosonic string theory, the Virasoro generators are 
written as
\begin{eqnarray}
L_0=-1+\frac{\nu^2}{2}-\frac{\hat \mu^2}{2}+\nu {\cal J}_L +\hat{\mu} {J}_L +
 N + \f{\al}{4}\vec{p}\,_{22}^{2} ~,\nonumber\\
{\ti L_0}=-1 +\frac{\nu^2}{2}-\frac{\hat \mu^2}{2}-\nu{\cal J}_R 
- \hat{\mu} {J}_R + {\ti N} + \f{\al}{4}\vec{p}\,_{22}^{2} ~,
\label{LL}
\end{eqnarray}
where 
$\hat{\mu}=(\mu-[\mu])$, while $[\mu]$ denotes 
the greatest integer less than or equal to $\mu$, and 
$\vec{p}_{22}$ abbreviates the momentum along the remaining ${\mathbb R}^{22}$. 
Taking the zero-slope limit, i.e. 
\begin{align}
\al\rightarrow0 ~,~~~
\beta\rightarrow0 ~,~~~\text{with}~~
b=\frac{\beta}{\al}~~\text{fixed} ~,
\end{align}
as well as defining 
\begin{align}
\begin{aligned}
\f{\al}{2}M^2:= \frac{1}{2} \big(\alpha^+_0 \alpha^-_0 + \alpha^-_0 \alpha^+_0 + 
\ti \alpha^+_0 \ti \alpha^-_0 + \ti \alpha^-_0 \ti \alpha^+_0 \big) 
- \f{\al}{2}\vec{p}\,_{22}^{2} ~,
\label{Nm}
\end{aligned}
\end{align}
we have the mass spectrum as%
\footnote{
We have 
assumed that as $\beta\rightarrow0$, $1>|\mu|>0$, so that $\mu= \hat \mu$.} 
\begin{align}
M^2= \f{2\mu}{\al}(l_L+ l_R +1) ~.
\label{ma}
\end{align}
The commutation relations $[ \alpha^+_0 , \alpha^-_0 ]= - i \nu$, 
$[ \tilde \alpha^+_0 , \tilde \alpha^-_0 ] = i \nu$ are used, and 
terms involving higher oscillating modes decouple, such 
as $(N +\ti N -2)$ and $\nu^2 - \hat \mu^2$. 
Here, 
$l_L(l_R)=0,1,2,...$ denotes the quasi-zero mode eigenvalue of 
$J_L(-J_R)$.

Next, we determine whether 
the mini-superspace approximation gives the same result. 
We are going to solve the wave equation $\triangle_0 \Psi=0$, where $\triangle_0$ 
is as in 
\eqref{L}. 
Assuming that the 
wave function takes the form 
\begin{align}
\Psi(t,r,\theta,\varphi,\vec{x})
=\Psi_t (t)\Psi_{r}(r)e^{ik\theta+im\varphi +i\vec{p}\vec{x}} ~,~~~~{m\in\mathbb Z} ~,
~~{k\in\mathbb R} ~,
\label{tw}
\end{align}
we find its temporal and radial parts, respectively, satisfy
\begin{eqnarray}
\Big[-\f{1}{t}\f{\p}{\p t} t\f{\p}{\p t} - \f{k^2}{t^2} 
- b^2 m^2 t^2\Big]\Psi_t (t)&=&{E}^2 \Psi_t (t) ~,\nonumber\\
\Big[-\f{1}{r}\frac{\p}{\p r} r\frac{\p}{\p r}+\Big(\f{m^2}{r^2}+b^2 k^2 r^2\Big)
\Big]\Psi_{r}(r)&=&R^2 \Psi_{r}(r) ~.
\label{ti}
\end{eqnarray}
The radial equation can be solved, and we find 
\begin{align}
\Psi_{r} (r)&=C r^{|m|} e^{\f{-b|k|r^2}{2}} L^{|m|}_{j} (b|k| r^2) ~,~~
\label{Con}
\end{align}
where $C$ is a constant. Also, here we have 

\begin{align}
E^2-\vec{p}\,^2_{22}=R^2=2b|k|\big(2j+|m|+1\big) ~,
\label{E}
\end{align}
where $j$ is a non-negative integer 
and 
$L^{|m|}_{j} (z)$ is the Laguerre polynomial, which is defined as 
\begin{align}
L^{m}_{j} (z) = \sum^{j}_{p=0} (-1)^{p}
 \left(
  \begin{array}{c}
   j+m \\
   j-p 
  \end{array}
 \right)
 \f{z^p}{p!} ~.
 \label{Lagu}
\end{align}
With the identifications 
\begin{align}
M^2=R^2 ~,~~~~m=l_L-l_R ~,~~~~~2j +|m|=l_L+ l_R ~,~~~~{\mu}={\al}bk ~,~~~~{\nu}={\al}bm ~,
\label{mll}
\end{align}
it is seen that Eq.~$\eqref{E}$ actually coincides with Eq.~$\eqref{ma}$.

In order to treat the temporal part of Eq.~$\eqref{ti}$, 
we carry out the change of variable $t\rightarrow 
e^{\xi}$ and $\Psi_t (t)\rightarrow { Y}(\xi)$ $(0\le t\le\infty)$. 
Then the differential equation becomes%
\footnote{Charged particles in the Rindler space 
obey a similar differential equation (see Ref.~17)).} 
\begin{align}
\Big(\f{d^2}{d\xi^2}+{E}^2 e^{2\xi}+b^2 m^2 e^{4\xi}+k^2\Big){Y}(\xi)=0 ~.
\label{tieq}
\end{align}
Next, setting $W(z)=e^{\xi}Y(\xi)$, where $z=-ib|m|e^{2\xi}$ is an imaginary variable, we arrive at 
\begin{align}
\Big[\f{d^2}{dz^2}+\Big(\frac{-1}{4} + \frac{\lambda}{z} +\frac{1-4\eta^2}{4z^2}\Big)\Big]W_{\lambda,\eta}(z)=0 ~,
&&\lambda=\frac{i}{4b|m|}{E}^2 ~,
&&\eta=\frac{i|k|}{2} ~.
\label{whi}
\end{align}
We thus find that $\Psi_t(t)$ 
can be expressed in terms of a Whittaker function. 
Finally, note that given an 
$x$-$p$ representation, as in Ref.~5) for quasi-zero modes, i.e. 
$\alpha ^{\pm}_0 = \big(i \sqrt{\frac{\alpha '}{2}} \partial_{\mp} \pm \frac{\nu} {\sqrt{2 \alpha '}} 
x^{\pm}\big)$, $\tilde \alpha ^{\pm}_0 =\big(i \sqrt{\frac{\alpha '}{2}} \partial_{\mp} \mp 
\frac{\nu}{\sqrt{2 \alpha '}} x^{\pm}\big)$, we 
can reproduce the first line in Eq.~$\eqref{ti}$ by plugging this 
representation into the LHS of the 
on-shell constraint $\eqref{ma}$. 

\subsection{Winding string production}

Because the model $\eqref{me2}$ is non-static, any quantum field propagating 
within it will give rise to particle production. By deriving the Bogoliubov 
coefficients, we can explicitly obtain the production rate.

Restricting ourselves to the time interval $0\le t\le\infty$ 
($-\infty\le\xi\le\infty$), 
we set the temporal part of the outgoing state to be 
$\psi_{\sigma}^{out}$=$e^{-\xi}W_{\lambda,\eta}(-z)$, 
where 
$\sigma$ represents the quantum number $(k,m,\vec{p})$ in Eq.~$\eqref{tw}$. Due to the property 
\begin{align}
z\rightarrow \infty ~, ~~~~~W_{\lambda,\eta}(-z)\simeq e^{z/2}(-z)^{\lambda} ~,
\end{align}
in the limit $t\rightarrow\infty$, we obtain 
\begin{align}
\psi_{\sigma}^{out}(t)\propto|t|^{2\lambda-1} e^{\f{-ib|m|t^2}{2}} ~;
\end{align}
i.e. it asymptotically approaches a positive frequency mode in the far future. 
Making use of the relation between the Kummer and the Whittaker functions 
\begin{align}
\begin{aligned}
W_{\lambda,\eta}(-z)&=\f{\G(-2\eta)}{\G(\f{1}{2}-\eta-\lambda)}e^{i\pi (\eta+\f{1}{2})}[M_{\lambda, -\eta}(-z)]^{\ast} +
\f{\G(2\eta)}{\G(\f{1}{2}+\eta-\lambda)} M_{\lambda, -\eta}(-z) ~,
\label{kumm}
\end{aligned}
\end{align}
where $M_{\lambda, \eta}(-z)=e^{i\pi (\eta+\f{1}{2})}[M_{\lambda, -\eta}(-z)]^{\ast}$, we are able to relate 
$\psi_{\sigma}^{out}(t)$ to the incoming state through 
the Bogoliubov transformation expressed by 
\begin{eqnarray}
\psi_{\sigma}^{in}&=&\alpha_{\sigma}\psi_{\sigma}^{out} + \beta_{\sigma}\psi_{-\sigma}^{out \ast} ~,\nonumber\\
\psi_{\sigma}^{out}&=&\alpha_{\sigma}^{\ast}\psi_{\sigma}^{in} - \beta_{\sigma}\psi_{-\sigma}^{in \ast} ~.
\label{bogol}
\end{eqnarray}
Thus, according to $\eqref{kumm}$, we can express the temporal part of the $in$ state in terms of
the Kummer function 
$\psi_{\sigma}^{in}=|t|^{-1} M_{\lambda, -\mu}(-z)$. 
Near $t=0$, it behaves as 
\begin{align}
\psi_{\sigma}^{in}(t) \propto e^{-i|{k}|\log|t|} ~,
\label{ins}
\end{align}
due to the property 
\begin{align}
z\rightarrow 0 ~, ~~~~~M_{\lambda, -\eta}(-z)=(-z)^{-\eta+\f{1}{2}}\big[1+{\cal O}(-z)\big] ~.
\label{due}
\end{align}
It is thus seen that the $in$ state contains both positive and negative 
frequency modes 
in the far future; 
i.e. winding strings are pair created. 
The production rate is found to be 
\begin{align}
|\frac{-\beta_{\sigma}}{\alpha^\ast_{\sigma}}|^2=|\gamma_{\sigma}|^2=e^{2\pi i\eta}
\Big|\frac{\Gamma(\f{1}{2}-\eta+\lambda)}{\Gamma(\f{1}{2}-\eta-\lambda)}\Big|^2 = 
\frac{1+e^{\pi\Big(|k|-\f{{ E}^2}{2b|m|}\Big)}
e^{-2\pi|k|}}{1+e^{\pi\Big(|k|-\f{{ E}^2}{2b|m|}\Big)}} ~.
\label{rate}
\end{align}

Defining $E'^2=E^2-2b|k||m|$, where $\vec{p}\,_{22}^{2}$ is ignored 
for simplicity, we are able to set 
$e^{\sum_{\sigma,j}\log|\gamma_{\sigma}|^2}=e^{-({\cal A}-{\cal B})}$ with 
\begin{eqnarray}
{\cal A}&=&\sum_{\sigma,j}\log\big[1+e^{\pi(|k|-\f{{ E}^2}{2b|m|})}\big] = 
\sum_{\sigma,j}\sum^{\infty}_{n=1}\frac{(-1)^{n+1}}{n} e^{\frac{-n\pi \al E'^2}{2|\nu|}}\nonumber\\
&=&-2{\text {Im}}\Big[\int_{0}^{\infty}\f{du}{u}{\text {Tr}}~e^{-\pi u {\cal H}}
\Big] ~.
\end{eqnarray}
Note that 
\begin{eqnarray}
{\text {Tr}}e^{-\pi u {\cal H}}&=&
\sum_{k,m,j} \int_{- \infty}^{\infty} d{E'}^2 
\rho_{k,m,j} ({E'}^2) e^{- \frac{1}{2}\pi u\al \Big[{E'}^2 + 2b|k|(2j+1)\Big]}\nonumber\\
&=&\sum_{k,m} \frac{1}{4i\sin(\pi|\nu| u)~\sin(-\pi{|\mu| iu})} ~,
\label{Tr}
\end{eqnarray}
where we have used the following formulas$:^{3)}$ 
\begin{eqnarray}
\frac{1}{2i\sin y}&=&\sum_{j=0}^{\infty}e^{-i(2j+1)y} ~,\nonumber\\
 \rho_{k,m,j} ({E'}^2 )&=&\f{i}{2b|m|} \log{\Lambda} + 
 \f{1}{2\pi} \frac{d}{d{E'}^2}\log
 \frac{\Gamma(\f{1}{2}-i\f{{ E'}^2}{4b|m|})}{\Gamma(\f{1}{2}+i\f{{ E'}^2}{4b|m|})} ~,
\end{eqnarray}
where $\Lambda$ is an infrared cutoff. 
That ${\cal A}$ is now explicitly recast into an imaginary part 
of a one-loop vacuum free energy reflects the optical theorem.

Next, we interpret the subtraction term ${\cal B}$. 
It can be shown that for 
$2b|k||m|\simeq E^2$ and $|k|\gg1$, we have 
${\cal B}\rightarrow\sum_{\sigma,j}e^{-2\pi |k|}$, 
which is just the production rate of untwisted strings, as can be seen by setting 
$b=0$ in Eq.~$\eqref{tieq}$.%
\footnote{Note that this situation resembles the process of 
open string pair creation in the half S-brane. Following 
Ref.~16), we can soon derive this production rate.} 
Therefore, comparing Eq.~$\eqref{rate}$ and the 
corresponding relation in the 
case of charged particle pair 
creation in the Rindler space,$^{17)}$ we observe that ${\cal A}$, 
which is induced by 
the 
background 
$B$-field in Eq.~$\eqref{me2}$, corresponds to the usual Schwinger effect, 
whereas ${\cal B}$, which 
is due to the Milne 
metric $(b=0)$, corresponds to the gravitational Unruh effect of Rindler dynamics. 
\subsection{Disk one-point functions}

We now compute classical disk one-point correlators of D-branes by using 
wave 
functions derived above. There are two kinds of D-branes we will consider, i.e. a 
D1-brane wrapping $X^{\pm}=\frac{1}{\sqrt{2}}te^{\pm\theta'}$ 
and a D-brane instanton wrapping 
$X=r e^{i\varphi'}$. 
Expanding the DBI action around the metric $\eqref{me2}$,$^{18),19)}$
\begin{align}
S^{\rm DBI}=S_0 + \int d^{p+1} \rho  ~\frac{\delta S}{\delta\Phi}\Big|_0 \delta\Phi + ... ~,
\end{align}
one can gauge fix the worldvolume coordinates to be parallel to 
the spacetime coordinates that D-branes wrap. 
Identifying the dilaton fluctuation $\delta\Phi$ as $\Psi(x)$ 
in Eq.~$\eqref{tw}$, we find that the one-point function can be read off as
\begin{align}
\langle \Psi\rangle _{\rm disk}=
\tau_p \int d^{p+1} x ~
e^{-\Phi}\sqrt{-{\text {det}}(G+B)}\Psi(x) ~, 
\label{disk}
\end{align}
where $\tau_p$ is the D$p$-brane tension.

For the D1-brane localized at $X=r_0 e^{i\varphi'_0}$, 
with ${\cal J}_L + {\cal J}_R =0$ ($k=0$), by plugging 
the metric $\eqref{me2}$ into Eq.~$\eqref{disk}$, we obtain
\begin{align}
\langle \Psi_{k=0,m,j}\rangle^{\rm D1}_{\rm disk}=\tau_1 \int d\theta 
dt|t|~C_{\sigma}\Psi_{t,\sigma}(t) \times(\text{other spatial parts}) ~.
\label{op}
\end{align}
Here, 
$C_{\sigma}=\frac{1}{\sqrt{|\alpha_{\sigma}|^2 - |\beta_{\sigma}|^2}}$ 
represents the normalization 
of $\Psi_{t,\sigma}(t)$, 
which is determined by requiring the unitary condition 
$|\f{\alpha_{\sigma}}{C_{\sigma}}|^2 - |\f{\beta_{\sigma}}{C_{\sigma}}|^2=1$ 
in Eq.~$\eqref{bogol}$. 
By choosing $\Psi_{t,\sigma}(t)=|t|^{-1} M_{\lambda,0}(-z)$, we obtain
\begin{align}
O_t=\int_{-\infty}^\infty 
dt|t|~C_{\sigma}\Psi_{t,\sigma}(t) =\frac{{\cal Q}}{2}\sqrt{\frac{-i}{b|m|}} ~,
\label{Ot}
\end{align}
up to a prefactor ${\cal Q}$ depending on the regularization procedure, 
which we do not
consider here. 
For the D-brane instanton localized at 
$X^{\pm}=\frac{1}{\sqrt{2}}t_0 e^{\pm\theta'_0}$, 
with ${J}_L + {J}_R =0$ ($m=0$), the part relevant to the $X$-plane is
\begin{align}
O_X=\int d\varphi dr r ~C \Psi_r(r)=\sqrt{\frac{b|k|}{\pi}}\int_0^\infty dx~ L^0_j (x)
 e^{-x/2}=2(-)^j \sqrt{\frac{\pi}{b|k|}} ~,~~~~x=b|k|r^2 ~,
 \label{Or}
\end{align}
where we have used Eqs.~$\eqref{Con}$ and $\eqref{Lagu}$, 
and the normalization $C =
 \sqrt{\frac{(b|k|)^{|m|+1} j!}{\pi (j+|m|)!}}$, as well as the
definition of the Gamma function, 
\begin{align}
\Gamma (l+1)=l!=\int_0^\infty dt~ t^l e^{-t} ~.
\end{align}
In the next section, we determine whether these classical results reproduce 
those from the quantum cylinder amplitude 
in the $\al\ll1$ limit.

\section{CFT description}

\subsection{Torus amplitude}

It would be illuminating to see whether the mini-superspace approach is consistent 
with the CFT computation. To 
carry out this, we first derive the torus amplitude 
\begin{align}
Z(\tau) = {\text {Tr}}~q^{L_0}\bar{q}^{{\ti L}_0} ~,~~~q = e^{2\pi i\tau} ~, ~~~
\tau = \tau_1 + i\tau_2 ~,
\label{Zt}
\end{align}
where 
$L_0$ and ${\ti L_0}$ 
are as in Eq.~$\eqref{LL}$. As in Ref.~10), we insert delta functions of the forms%
\footnote{Note that the sum ${\cal J}_{L} + {\cal J}_{R}$ 
(or $J_{L} + J_R)$ is some real value.}
\begin{eqnarray}
&&1=\int d^2{j_b} ~\delta({\cal J}_L - j_b)\delta({\cal J}_R - \bar{j}_b) ~,\nonumber\\
&&\delta({\cal J}_L - j_b)\delta({\cal J}_R - \bar{j}_b)=
\int d^2\chi_b ~e^{2\pi i\chi_b ({\cal J}_L - j_b)
 +2\pi i\bar{\chi}_b{({\cal J}_R - \bar{j}_b)}} ~,\nonumber\\
&&1=\int d^2{j}_r ~\delta({J}_L - j_r)\delta({J}_R - \bar{j}_r)
~,\nonumber\\
&&\delta({J}_L - j_r)\delta({J}_R - \bar{j}_r)=
\int d^2\chi_r ~e^{2\pi i\chi_r ({J}_L - j_r)
 +2\pi i\bar{\chi}_r({J}_R - \bar{j}_r)} ~,
\label{del}
\end{eqnarray}
where $\delta^2(z)=\int d^2 y~ e^{2\pi iyz + 2\pi i\bar{y}\bar{z}}$, into $Z(\tau)$. 
This yields 
\begin{align}
\begin{aligned}[b]
Z(\tau)=\f{V_{22}}{(2\pi)^{22}(\al\tau_2)^{{11}}}
&\int d^2\chi_r d^2{\chi_b}~
\frac{1}{{\big| \vartheta_1(i\chi_b |\tau)\vartheta_1(\chi_r |\tau) \eta(\tau)^{18} \big|^2}}\\
&\hspace*{-3cm}\times
\Big[\int d^2 j_r d^2{ j_b}~ q^{-\nu j_b}q^{-(\mu-[\mu]) j_r} 
\bar{q}^{\nu \bar j_b} \bar{q}^{(\mu-[\mu]) \bar j_r}
e^{-2\pi i \big( \chi_b j_b +\bar{\chi}_b \bar j_b + \chi_r j_r + 
\bar\chi_r \bar j_r \big)} \Big] ~\\
&\hspace*{-3cm}\times
\exp\Big[\pi\frac{( \chi_b-\bar{\chi}_b )^2 -( \chi_r-\bar{\chi}_r )^2}{2\tau_2}\Big] ~.
\label{tor}
\end{aligned}
\end{align}
Note that terms involving $\nu^2$ and $\hat \mu^2$ in \eqref{Zt} 
have been summarized into the third line.%
\footnote{With the expression 
\begin{align}
\begin{aligned}
e^{2\pi \tau_2 (i\nu)^2} = \sqrt{\f{\tau_2}{2}}\int^\infty_{-\infty}d l 
~e^{\f{-\pi\tau_2}{2}l^2} 
e^{2\pi i\tau_2\nu l} ~,
\end{aligned}
\end{align}
we can shift $j_b \rightarrow j_b + \f{il}{2}$, $\bar{j}_b \rightarrow \bar{j}_b - 
\f{il}{2}$ in the second line of Eq.~\eqref{tor} 
to eliminate $e^{2\pi i\tau_2 \nu l}$, and then 
integrate out $l$ to obtain the third line. 
A similar manipulation can be performed to $e^{2\pi \tau_2 \hat \mu^2}$.} 
In order
to treat the second line above, 
we shift $\chi_r \rightarrow \chi_r + 
\tau[{\mu}]$ and $\bar{\chi}_r \rightarrow \bar{\chi}_r + \bar{\tau}[{\mu}]$. 
By doing this, the second line can be integrated out, and it becomes 
$\frac{1}{(2\beta\tau_2)^2}\exp\big[\frac{\pi}{\beta\tau_2}
( \chi_b\bar{\chi}_{r}-\chi_r\bar{\chi}_b )\big]$. Note also that we have used%
\footnote{It is proven in Ref.~15) that no negative norm 
states propagate within this kind of Lorentzian torus amplitude.}
\begin{align}
\begin{aligned}
{\text{Tr}} ~\big[q^{N-1}e^{2\pi i\chi_b {\cal J}_L +2 \pi i\chi_r J_L}\big]
{\text {Tr}} ~\big[\bar{q}^{\ti N- 1} e^{2\pi i\bar{\chi}_b {\cal J}_R +2 \pi i\bar{\chi}_r J_R}\big]
=\frac{1}{{\big| \vartheta_1(i\chi_b |\tau)\vartheta_1(\chi_r |\tau) \eta(\tau)^{18} \big|^2}} ~.
\label{trace}
\end{aligned}
\end{align}
Modular invariance becomes manifest when we simultaneously rescale
$\tau\rightarrow\frac{-1}{\tau}$, $\chi_{r,b}\rightarrow\frac{\chi_{r,b}}{\tau}$ for 
$\int\f{d^2\tau}{\tau_2}Z(\tau)$. 
Then, by employing the point-particle approximation, in which $\tau_1=0$, $\tau_2=u$
 and $\al\ll1$, only Eq.~$\eqref{trace}$ remains in the integrand of 
Eq.~$\eqref{tor}$. 
The (anti-) holomorphic quasi-zero mode contribution becomes 
\begin{align}
\begin{aligned}
\frac{1}{4\sin(\pi \nu u)~\sin(-\pi{\mu iu})} ~,
\label{36}
\end{aligned}
\end{align}where we have renamed 
$\chi_b$ as $-\nu\tau$ and $\chi_r$ as $-\hat{\mu}\tau$, 
comparing 
the trace structure of Eq.~$\eqref{trace}$ with Eq.~$\eqref{LL}$. 
It is obvious that Eq.~$\eqref{36}$ is reproduced by Eq.~$\eqref{Tr}$.

The double-pole 
structure in Eq.~$\eqref{trace}$, which has no imaginary part 
after carrying out the moduli integral, implies that there is 
no particle production, according to the optical theorem. 
We can understand this by 
considering the 
fact that in CFT the $in$ and $out$ vacua coincide, because the 
physical descriptions are the same at 
$t=\pm\infty$, as seen from the metric $\eqref{me2}$. 
By contrast, in the mini-superspace approach, we 
have chosen the $in$ and $out$ vacua to be at $t= 0$ and $t = \infty$, respectively. 
Hence, there is no 
contradiction with the observation concerning string production made previously. 

\subsection{Cylinder amplitude}

We now read off the disk one-point function from 
the cylinder amplitude and take the zero-slope limit 
to see if it does match with the classical result. 
We first treat the case 
of a D1-brane which wraps $X^{\pm}$, 
but moves along the $\varphi'$ direction of $X$ at $r_0$ with an angular 
momentum $m$. From the arguments made in deriving Eq.~$\eqref{pe}$, 
it is seen that $\beta m$ 
is the twist 
parameter of the $X^{\pm}$-plane. The cylinder amplitude is found to be%
\footnote{An overall volume factor is omitted here.}
\begin{eqnarray}
&&\frac{\alpha' \pi}{2} {\cal N}_1^{'2}  \int_0^{\infty}  ds~
\langle\langle B,m;D1|
e^{- \pi s (L_0 + \tilde L_0)}|B,m;D1\rangle\rangle\nonumber\\
&&\hspace*{1cm}=\int_0^{\infty} ds
    \frac{ (\f{{\cal N}_1^{'}}{{\cal N}_1})^2 4\pi^2 \al
     e^{- \pi \beta^2 m^2 s-\frac{\pi  m^2 \al}{2r_0^2 }s}}
   {s^{\frac{23}{2}}(8 \pi^2 \alpha')^{\frac{3}{2}}\vartheta_1 
   (\beta|m|s{\big|}is)\eta(is)^{21}} ~,
\end{eqnarray}
where ${\cal N}_1^{'} = \sqrt{\frac{8 \pi^2 \alpha'}{2 \sinh ( \pi \beta|m|)}}
{\cal N}_1$ 
and ${\cal N}_{1}$ 
is the usual D1-brane normalization. From the 
structure of the above expression, we see that for 
${\al\ll1}$, $(\f{{\cal N}_1^{'}}{{\cal N}_1})^2 \propto 
|O_t|^2 $ of Eq.~$\eqref{Ot}$. 
There remains a proportionality ambiguity 
which involves the regularization procedure mentioned below Eq.~$\eqref{Ot}$. 
We next calculate the overlap of two D-brane instantons $|B,k;Ins\rangle\rangle$. 
Because the D-brane instanton wraps 
$X$ at 
$t_0$, with a momentum $k$ along the $\theta'$ direction of $X^{\pm}$, 
we have (for $\beta |k|<1$) 
\begin{eqnarray}
&&\frac{\alpha' \pi}{2}N_1^{'2}  \int_0^{\infty}  ds~
\langle\langle B,k;Ins|
e^{- \pi s (L_0 + \tilde L_0)}|B,k;Ins\rangle\rangle\nonumber\\
&&\hspace*{1cm}=\int_0^{\infty} ds
    \frac{(\f{N_1^{'}}{{ \cal N}_{1}})^2 4\pi^2 \al 
    e^{\pi \beta^2 k^2 s-\frac{\pi    k^2 \al}{2t^2_0 }s}}
   {  s^{\frac{23}{2}}(8 \pi^2 \alpha')^{\frac{3}{2}}\vartheta_1 
   (-i \beta|k|s{\big|}is)\eta(is)^{21}} ~,
\end{eqnarray}
where 
$N^{'}_1 = \sqrt{\frac{8 \pi^2 \alpha'}{2 \sin(\pi \beta|k|) }}{\cal N}_{1}$. 
In the limit ${\al\ll1}$, 
we find that $(\f{N^{'}_1}{{\cal N}_1})^2$ exactly 
coincides with the square of $O_X$ derived in 
Eq.~$\eqref{Or}$.

\section*{Acknowledgements}

We are grateful to Y. Hikida, F.-L. Lin, Y. Matsuo, Y. Sugawara, Y. Tachikawa 
and, especially, 
T. Takayanagi for stimulating discussions.

\end{document}